\documentclass[aps,prl,twocolumn,superscriptaddress,floatfix]{revtex4-2}
\usepackage{graphicx,amssymb,amsfonts,amsmath,chemarr,color,commath}
\usepackage[dvipsnames]{xcolor}
\usepackage{soul}

\newcommand{\awesom}{\textbf{\texttt{aweSOM}} }
\newcommand{\tristan}{\texttt{Tristan-mp} }

\newcommand{\dpvi}{$d_{\rm PVI}$}

\begin{document}

\title{Emergence and Detection of Electron-Scale Current Sheets in Turbulence with MMS Observations and fully kinetic 3D simulations}





\author{Zachary Davis}
\email{zkdavis@hawaii.edu}
\affiliation{Institute for Astronomy, University of Hawaii, Manoa, 2680 Woodlawn Dr., Honolulu, HI 96822, USA}

\author{Alexandros Chasepis}
\affiliation{Laboratory of Physics and Chemistry of the Environment and Space (LPC2E), OSUC, Univ Orleans, CNRS, CNES, 45071 Orleans, France}
\affiliation{Laboratory for Atmospheric and Space Physics, University of Colorado Boulder, Boulder, CO 80303, USA}

\author{Colby Haggerty}
\affiliation{Institute for Astronomy, University of Hawaii, Manoa, 2680 Woodlawn Dr., Honolulu, HI 96822, USA}
\affiliation{Department of Physics and Astronomy, University of Iowa, Iowa City, IA 54224, USA}

\author{Luca Comisso}
\affiliation{Department of Astronomy and Department of Physics, Columbia University, 538 West 120th Street, New York, NY 10027, USA} 

\author{Derek Sikorski}
\affiliation{Institute for Astronomy, University of Hawaii, Manoa, 2680 Woodlawn Dr., Honolulu, HI 96822, USA}
\begin{abstract}






The solar wind is characterized by turbulence, where a cascade produces intermittent current structures called current sheets (CS) that efficiently dissipate energy into the plasma. These have been studied with in situ spacecraft observations, but single-spacecraft techniques such as the partial variance of increments (PVI) are inherently limited since they lack spatial context. A combined analysis of in situ observations and numerical simulations can provide significant insight into the properties of intermittent structures forming in heliospheric turbulence. Understanding the size and distribution of these structures is crucial in tracing the pathways of energy dissipation and particle energization in space plasma. Using 3D fully kinetic simulations of magnetized turbulence, we identify CS via machine learning and find a complex broken-power-law distribution for the CS widths, where the power-law breaks separate ion-scale CS from electron-scale CS. Electron-scale CS dominate, with widths peaking near $2d_e$. Comparing simulations with MMS data, we test PVI as a CS detector and show it can infer CS scale, though oblique crossings inflate inferred sizes. The prevalence of electron-scale sheets suggests they may contribute to plasma heating in aggregate.

\end{abstract}

\maketitle

\textit{Introduction}:
The solar wind, a weakly collisional, magnetized outflow from the Sun, offers a rare opportunity to measure fundamental plasma processes directly. For many astrophysical environments, we are limited to inferring plasma properties from the electromagnetic radiation that reaches our telescopes, but in the case of near-Earth plasmas, we have the unique opportunity to fly spacecraft directly into the medium to measure plasma properties {\it in situ} \citep{BrunoCarboneLRSP13}. The solar wind spans a wide range of spatial scales, separating where energy initially enters the system from where it will eventually be dissipated. At large Reynolds numbers, turbulence mediates this energy transfer through a cascade where initially large fluctuations, via non-linear interactions, transfer energy to smaller scales until reaching a scale where efficient dissipation can take place. During this cascade, large fluctuations break apart into smaller, more intermittent structures that are stretched along the magnetic field, creating large anisotropy in the plasma \citep{goldreich1995}. Normal to the magnetic field, the structures thin throughout the cascade and often have large magnetic field gradients that produce a current, commonly referred to as current sheets (CS). Magnetic reconnection within CS provides a mechanism for converting magnetic energy into thermal and non-thermal particle energy \citep{French2023}, and recent particle-in-cell (PIC) simulations of magnetized turbulence have shown that particles are preferentially injected into the non-thermal population near or within CS \citep{comisso2018,comisso2019,Joonas2021}. However, CS in turbulence are highly intermittent, making dissipation and acceleration regions in the plasma highly localized, taking up only a small fraction of the volume \citep{Zhdankin2013,zhdankin2016,wan2016,Davis2024}.

CS have already been individually identified in the solar wind \citep{Phan2006,Gosling2007,Eriksson2016}, but building a statistical picture from a single spacecraft trajectory typically relies on the partial variance of increments (PVI) \citep{Greco2018SSR, Greco2008,Greco2009,Servidio2011}, which has limited spatial context for determining whether a spike corresponds to a CS crossing or for accurately encoding the structure size \citep{Greco2016}. PVI has been used to characterize turbulent intermittency across a range of environments, including the inner heliosphere \citep{Chhiber2020} and Earth's magnetosheath \citep{Chasapis2017ApJ}. Multi-point observations at kinetic scales allow us to overcome some of these limitations by directly measuring the spatial variations at a scale fixed by the spacecraft separation. The MMS mission \citep{Burch2016} revealed an increased intermittency at sub-ion kinetic scales \citep{Chhiber2018JGR}, and an abundance of current sheets at those scales \citep{Yordanova2016,Yordanova2020}, along with evidence that CS may provide a significant pathway for energy dissipation \citep{Chasapis2018ApJL}. Observations of Earth's magnetosheath have further shown that sub-ion scale CS can exist decoupled from the ion scale physics with extremely localized heating from smaller electron scale CS and that reconnection can occur within these small-scale sheets \citep{Phan2018,Stawarz2022}. Although individually, these events may have little effect on the plasma temperature, their cumulative contribution to the local heating rates may be important, not unlike nanoflare theories of coronal heating \citep{Parker1988}. 

However, the limited spatial coverage of \emph{in situ} spacecraft trajectories constrain us from obtaining a fuller picture of the structure and properties of CS populations forming in turbulence. Past numerical studies have attempted to address this by comparing observational measures such as the PVI with results obtained from predominantly 2D or non-fully-kinetic simulations \citep{Finelli2021,Wan2015}. Multi-point observational results have demonstrated the highly anisotropic nature of turbulent structures \citep{Chasapis2020}, as well as the predominance of structures forming at electron-scales \citep{Chhiber2018JGR,Stawarz2022}. However, without validation against three-dimensional fully kinetic simulations, it remains unclear whether CS size distributions inferred from PVI reflect the true underlying population. PIC simulations of magnetized turbulence can directly address this, providing 3D kinetic-scale measurements of CS as ground truth against which PVI can be tested.

In this letter, we investigate the link between CS in PIC simulations and solar wind observations by first directly analyzing the distribution of CS sizes in simulation, before validating how effective PVI is at detecting CS in 3D simulations and comparing with MMS data. We then measure the distance traveled inside a CS by a simulated spacecraft and compare it with the distance inferred from PVI to test how reliably PVI can infer CS crossing scales, before examining the distribution of sizes inferred from MMS measurements.

\textit{Methods}: In this work, we analyze the fully kinetic PIC simulation of ion-electron turbulent plasma originally performed in \citet{Comisso2022}.  In \citet{Comisso2022} the Vlasov-Maxwell system of equations is solved using the particle-in-cell (PIC) method with \tristan \citep{Spitkovsky2005}. The simulation domain is a triply periodic cube with  side lengths $L=1400$ cells or $L\approx 424.24 d_{e0}$. $d_{e0}= c / \omega_{pe0}$ is the electron inertial length, defined by the electron plasma frequency $\omega_{pe0} =\sqrt{4\pi n_0 e^2 /m_e}$, and is related to the spatial resolution by $\Delta x = d_{e0}/3.3 $. $n_0$ is the plasma density. The simulation time step is $\Delta t \approx 0.136 \omega_{pe0}^{-1}$. Each cell in the simulation contains an average of 64 particles. 

The plasma is an ion-electron plasma with each species initially distributed as a Maxwellian. Ions have a reduced mass of $m_i = 50 m_e$. 
The plasma is initialized with a uniform background magnetic field $\mathbf{B_0} = B_0 \mathbf{\hat{z}}$ and turbulence is initialized via magnetic fluctuations with transverse polarizations to $\mathbf{B_0}$ (see \citet{Comisso2022,comisso2018,comisso2019} for additional details). The fluctuation amplitude is set so that $\delta B_0/B_0 = 1$, with $\delta B_0 = \langle \delta B^2(t=0)\rangle^{1/2}$, and is injected with the coherence scale $l_0 = L/3$. The plasma has an initial total $\beta_0 = \beta_{i0} + \beta_{e0} = 0.32$. Results use normalized units: magnetic field $\mathbf{b} = \mathbf{B}/B_0$, current density $\mathbf{j} = \mathbf{J}/ e n_0 c$, and fluid velocity $\mathbf{v} = \mathbf{V}/c$. We analyze the simulation once $\langle j^2\rangle$ has peaked at $t=1.25  l_0/v_A$. At this point turbulent dissipation has peaked and the turbulent cascade is fully developed.

To identify and characterize CS we apply the self-organizing-map (SOM) method of \citet{Davis2026}, implemented with \awesom \citep{Ha2025}, here trained on the signed current density $\mathbf{j}$ across all time steps. Following \citet{Davis2026}, the current density is normalized to twice its root mean square value and capped at unity, a threshold consistent with prior CS studies \citep{Zhdankin2013,wan2016}, and the SOM clusters are combined to match the number of cells above this threshold, yielding masks for $j>0$ and $j<0$. Connected structures are identified from each mask and sliced along the guide field, with each connected piece within a slice treated as a separate segment. For each segment a spline fit gives the perpendicular arc length $l_\perp$, and walking along the local normals gives the width $w$. From these we form the aspect ratio $\alpha=l_\perp/w$ and retain segments with $\alpha\ge1$. We discard structures and segments below $90 d_e^3$ and $1d_e^2$ respectively to limit small-sample error. All SOM training parameters and the segmentation, spline-fitting, and width-measurement procedures are identical to those used in \citet{Davis2026}.


To compare with the simulation we use observations by the MMS spacecraft in the Earth's magnetosheath, focusing on an interval of burst-resolution data on December 26 2017, from 06:12:43 to 06:52:23. The interval is 40 minutes long, which corresponds to several times the typical correlation scale in the magnetosheath \citep{Stawarz2022}, providing us with a sufficient statistical sample to study the properties of intermittent structures. The proton gyro-radius is $\rho_i \sim 101 km$ and the ion inertial length is $d_i \sim 48 km$, while for the electrons respectively $\rho_e \sim 1 km$ and the electron inertial length is $d_e \sim 1 km$ .The spacecraft separation is $18.9 km$, which allows us for a direct calculation of two-point increments below ion kinetic scales. 
We use burst-resolution FGM magnetic field data \citep{Russell2016} to compute the PVI, and plasma moments from the FPI instruments \citep{Pollock2016} to calculate the background velocity and the relevant plasma parameters.  

\begin{figure}[ht]
    \centering
    \includegraphics[width=1\linewidth]{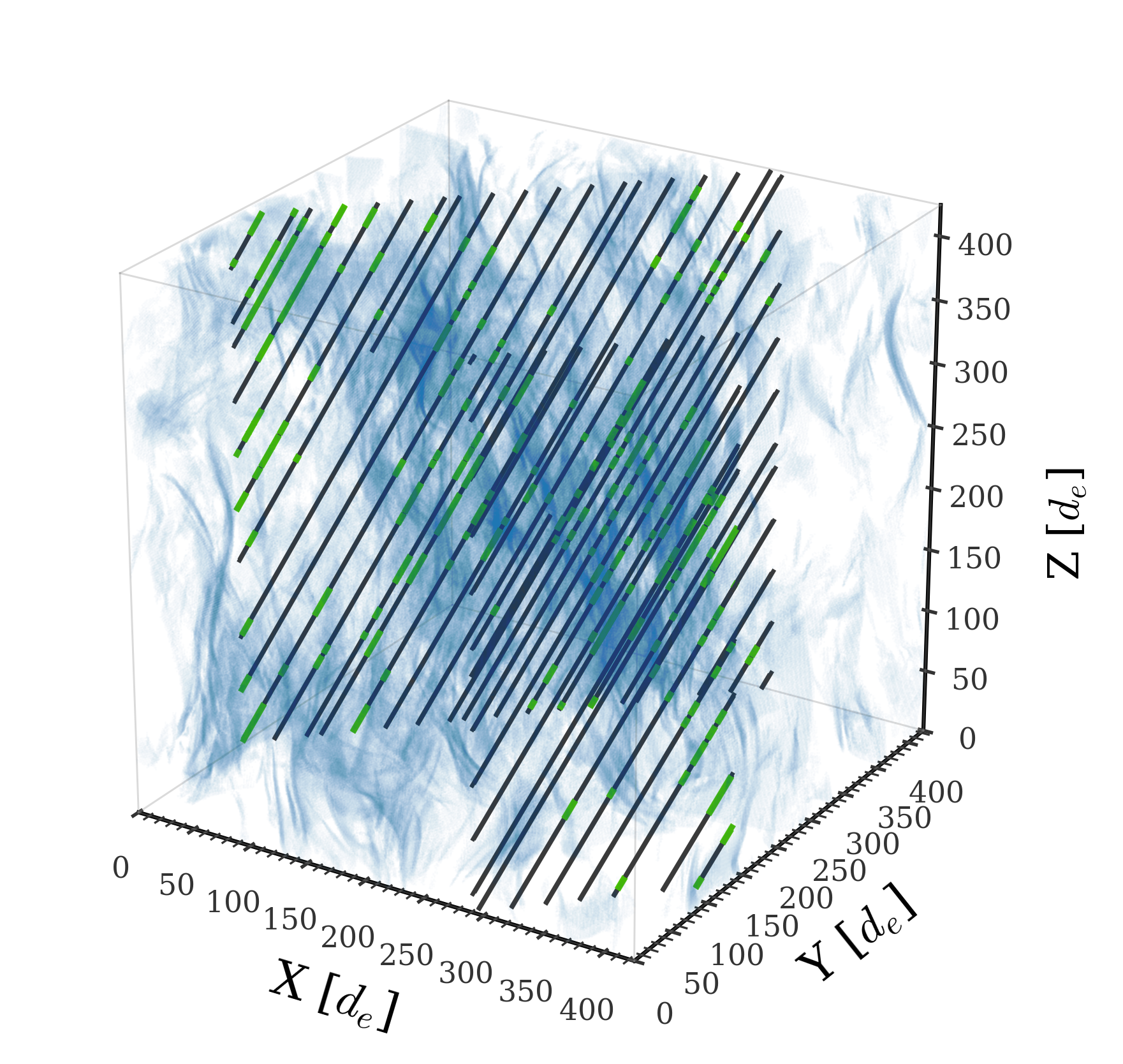}
    \caption{A 3D visualization of the simulation at peak turbulent dissipation, with CS shown in blue. The simulated spacecraft trajectory (black line) is marked with green points where it intersects an identified current sheet.}
    \label{fig:3d_trajectory}
\end{figure}

In order to make a direct comparison in the simulation we use PVI with methods outlined in \citet{Greco2018SSR}. Our trajectory is meant to mimic the MMS trajectory by initially starting in a fixed location and then traveling at a 45 degree azimuthal angle in the $x$-$y$ plane and a 60 degree elevation angle above the $x$-$y$ plane. This gives a similar angle between the trajectory and the background magnetic field as the MMS data. It then continues in a straight line exiting and entering the box periodically. The path is illustrated in Figure \ref{fig:3d_trajectory}. Then along the path we measure the spatial magnetic variance $\Delta b = b(r + \Delta r) - b(r)$, where $\Delta r$ is the separation between a pair of such paths, mimicking the two-spacecraft MMS measurement, and compute the normalized magnitude,
\begin{equation}
\label{eq:PVI}
PVI = \frac{\Delta b}{\sqrt{\langle \Delta b ^2 \rangle}}.
\end{equation}
For a point $r$ not on the simulation grid, we use cubic interpolation to find the value of the field at that point. The average of $\langle \Delta b^2 \rangle$ is over the entire path length of $30L$. We choose the separation $\Delta r$ to be about 1.12 $d_e$, the MMS spacecraft separation rescaled by the ratio of ion kinetic scales in the two systems ($\rho_i^{sim}/d_i^{MMS}$), and sample the path every 1.68 $d_e$ to match the MMS cadence.

\begin{figure}[ht]
    \centering
    \includegraphics[width=1\linewidth]{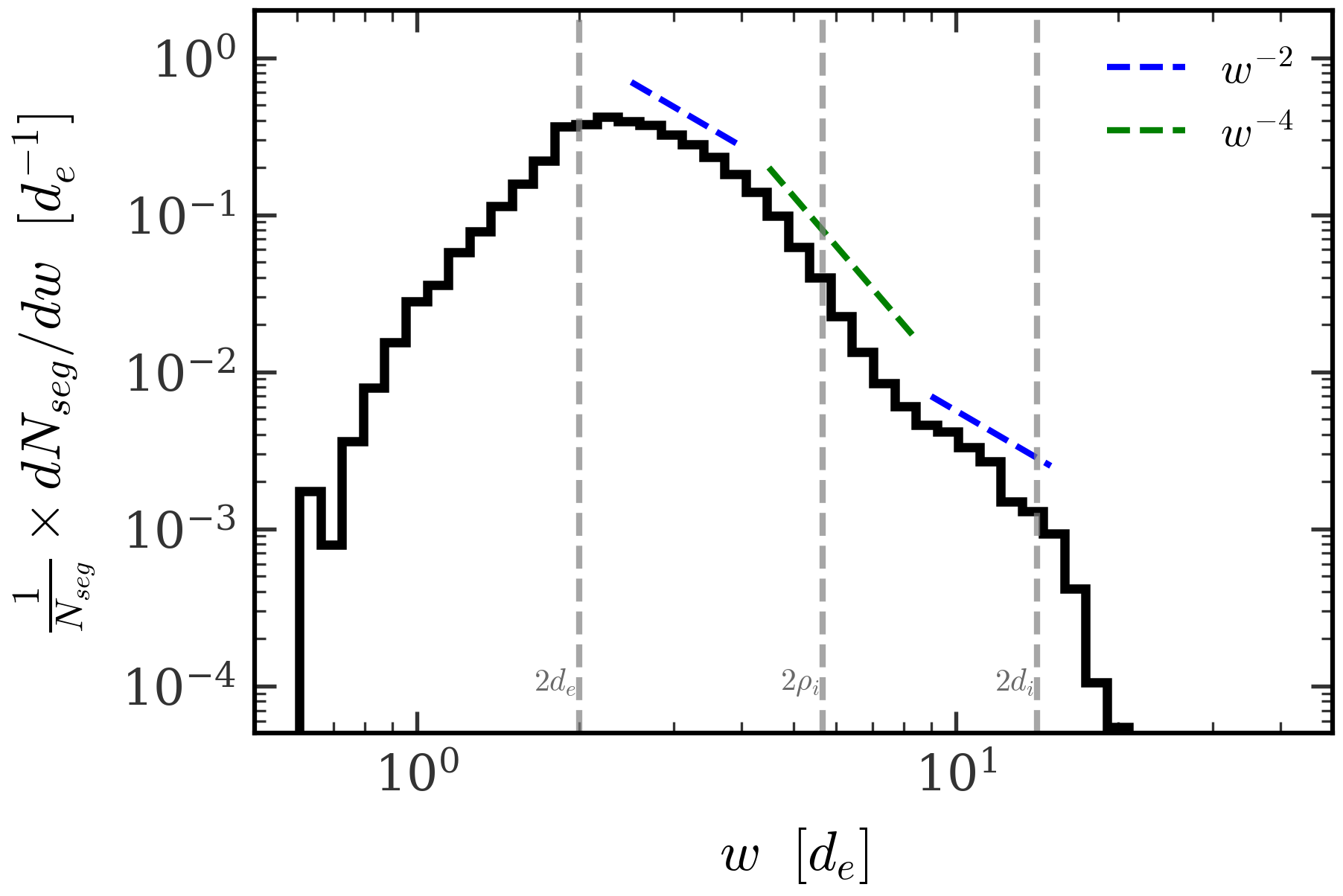}
    \caption{PDF of directly measured current sheet widths $w$ for all segments. Vertical dashed lines indicate $d_e$, $\rho_i$, and $d_i$, with dashed colored lines showing power-law fits to each scaling range and indices given in the legend.}
    \label{fig:width_pdf}
\end{figure}

\textit{Results}: Figure \ref{fig:width_pdf} shows the PDFs of the measurements $w$ from the simulation, which peak near $2d_e$ and follow a broken power-law separated by their physical scales.. Electron scales dominate near the peak with a power-law index of -2 but drop off as they approach the ion-scale gyro-radius ($\rho_i$) changing to an index of $\sim -4 $ before flattening again to $-2$ until a little above $2 d_i$.  Overall, electron scales dominate the distribution of $w$ measurements, making it more likely for a random crossing to interact with structures in the electron-scale range than in the ion-scale range. The dominance of electron scales suggests that many sheets in this simulation lie in the scale range where electron-only reconnection can occur, similar to events observed in Earth's magnetosheath \citep{Phan2018}, where ions remain decoupled from the reconnecting field when $w \ll d_i$ \citep{Sharma2019,Boldyrev2019,Vega2020,Vega2023}.

\begin{figure}[ht]
    \centering
    \includegraphics[width=1\linewidth]{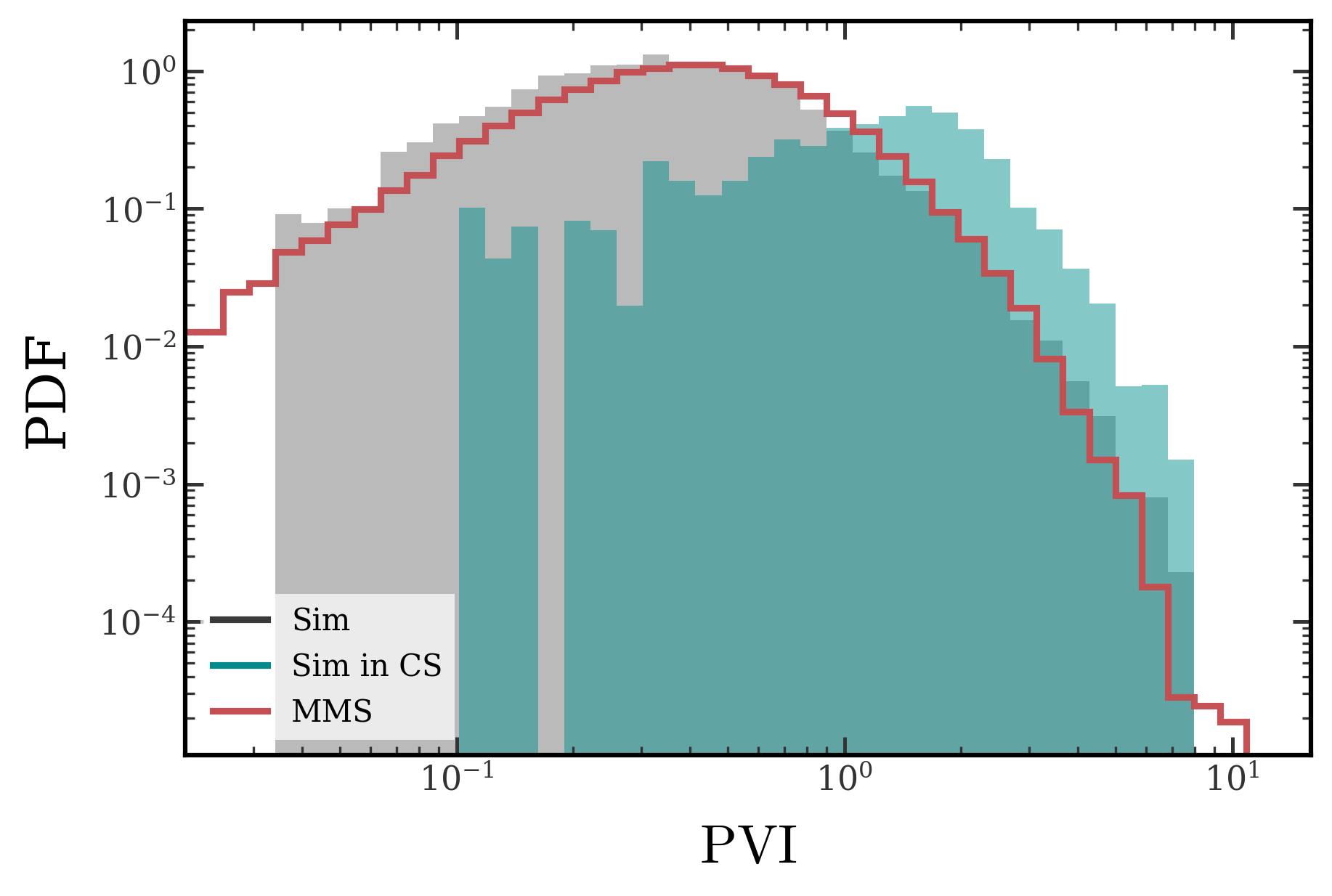}
    \caption{PDFs of $PVI$ for the full simulated trajectory (grey), points on the trajectory confirmed to lie inside a current sheet in the simulation (teal), and MMS observations (red). }
    \label{fig:PVI_pdf}
\end{figure}

\begin{figure}[ht]
    \centering
    \includegraphics[width=1\linewidth]{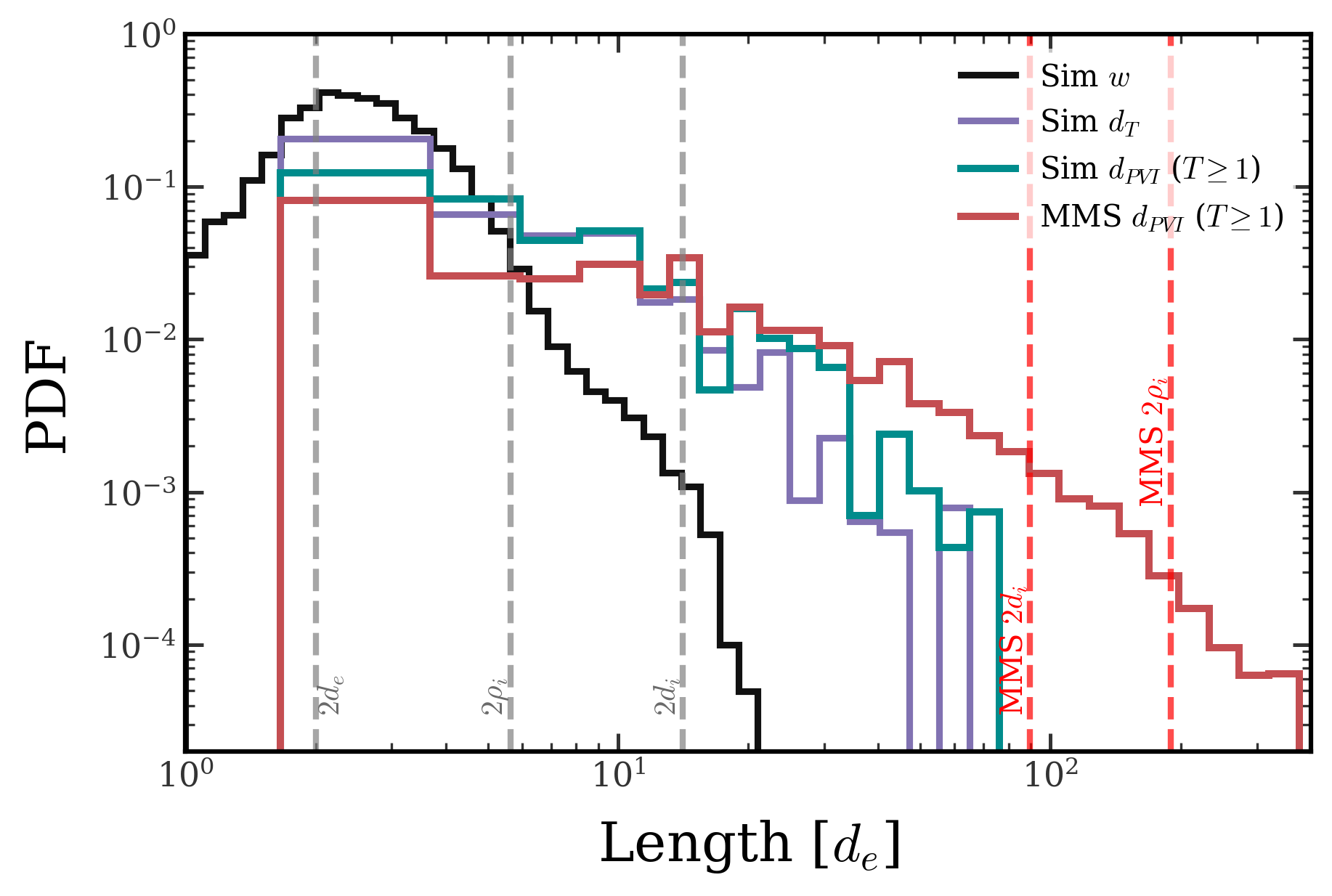}
    \caption{PDFs of current sheet size estimates from the simulation and MMS. The direct path length $d_T$ (purple) serves as the ground truth, with PVI-inferred distances \dpvi{} shown for the simulation (teal) and MMS (red), and $w$ (solid black) included for reference. Bins for purple, teal, and red are merged at small scales so that none are narrower than the 1.68 $d_e$ sampling interval. Vertical dashed lines mark twice the characteristic ion scales of the simulation (grey) and MMS (red).}
    \label{fig:pvi_width}
\end{figure}

The ability for PVI to detect CS in 3D is illustrated in Figure \ref{fig:PVI_pdf} where PVI from the simulation is compared with the MMS PVI and appears qualitatively similar. Additionally, we show the PVI measurements from the simulation restricted to trajectory points that lie within a CS, where a point is considered within a CS if either corresponding point on the dual path lies within a CS. Figure \ref{fig:PVI_pdf}  shows a clear shift in the PDF towards higher values in the PVI when inside of a current sheet. Although this indicates that PVI can be useful for detecting CS, we save the details concerning the optimal PVI threshold for a follow-up paper. 
To explore how well PVI can infer CS crossing scales, we introduce two new measurements in Figure \ref{fig:pvi_width}. First, we measure the exact distance traveled by the simulated trajectory from the point where it enters a CS to the point where it exits, referred to here as the path length distance $d_T$. $d_T$ serves as a reference for the best possible measurement that could be made with PVI given the sampling rate. This is the most direct comparison for PVI because both $d_T$ and \dpvi{} are measured along the same trajectory. We compare $d_T$ with \dpvi, the distance along the trajectory between the PVI measurement rising above a PVI threshold and falling below that threshold. For this work, we use a threshold of 1. We show PDFs for \dpvi{} for both the MMS data and the simulation for comparison. Finally, we show $w$, giving a comparison between the intrinsic width and the size experienced along the trajectory. Figure \ref{fig:pvi_width} shows that simulation measurements for \dpvi{} have qualitative agreement with the distribution from $d_T$. Further, it shows that if MMS data illustrates a similar scenario, it may be due to a distribution of structures spanning across kinetic scales but dominated by electron scales. 
Sensitivity of these results on the PVI threshold will be explored and discussed in a more extensive forthcoming manuscript.
The increase in overall size seen in Figure \ref{fig:pvi_width} compared with $w$ is due in part to angled paths through the CS rather than the minimum-width path used in the $w$ measurement. In general, using PVI to interpret CS size is consistent with $d_T$, showing that PVI could be used to infer structure size and, if so, may suggest that the solar wind includes a wide range of scales including sub-ion scales.

\textit{Discussion}: In this work, we use 3D fully kinetic PIC simulations of plasma turbulence relevant to the solar wind in order to characterize the statistical distribution of CS in the simulation and validate PVI as a method for detecting and characterizing the size of CS structures against MMS observations in Earth's magnetosheath. Current sheets are identified using Self-Organizing Maps, a machine learning technique. Using techniques developed in \citet{Davis2026} we analyze the statistical widths, $w$, of all CS in the simulation to compare characteristic measurement to what can be measured with PVI, which is in turn compared with the PVI obtained spacecraft measurements. The $w$ measurements in Figure \ref{fig:width_pdf} were peaked at $2d_e$ similar to  the relativistic results in \citet{Davis2026} but differ in showing a complex broken power-law distribution rather than an exponential decay. The power-law at peak had an index of $\sim-2$ before declining to $\sim-4$ while transitioning to ion scales before hardening again at $\sim-2$ up to a little beyond $2d_i$. This suggests a more complex distribution of CS structures when multiple particle species play a role in the plasma. 

We additionally use the segmented CS to test whether a PVI threshold can detect CS. The overall PVI distribution from the simulation resembles that from the MMS data, while trajectory points confirmed to lie inside a CS show a clear shift toward larger PVI values. This suggests that a threshold can help distinguish CS from the background plasma. However, we leave detailed testing and tuning of the preferred threshold value to a follow-up study. Furthermore, in Figure \ref{fig:pvi_width} we find that PVI can be used to infer trajectory scale by comparing the direct path length $d_T$ with the PVI-inferred distance \dpvi{}, which shows strong qualitative agreement with the $d_T$ measurement. However, since the trajectories pass through CS at an angle, the inferred scale is skewed towards larger values and possibly smooths out the broken power-law initially observed in the $w$ pdf to a single power-law extending to larger values and peaking after $2 d_e$. It is worth noting that these broad widths sampled in the simulation are similar to the \dpvi{} measurements from the MMS probe, suggesting that CS in the solar wind may also be composed of a complex, scale-dependent structure peaked at electron scales. Additionally, the broader scale separation for the plasma observed in the MMS measurements may also explain the extended distribution compared to the simulation results.

These findings have significant implications for understanding energy dissipation in solar wind and other astrophysical plasmas. The dominance of electron-scale CS suggests that these structures may provide a non-negligible contribution to energy conversion and particle energization, with a substantial part of turbulent dissipation occurring at kinetic scales below $\rho_i$. However, in different plasma turbulence regimes, dissipation may still be primarily mediated by ion-scale processes, and further work is needed to understand under which conditions such electron-scale structure formation is favored. Previous works have shown that on scales below $\rho_i$, reconnection-affected cascades can produce electron-only current sheets where the ions are completely decoupled from the structure \citep{Sharma2019,Vega2020,Vega2023,Phan2018}. The scales of the sub-ion structures may be limited by the tearing rate and the eddy turnover time \citep{Boldyrev2019}. Initial results from electron-only reconnection sites in Earth's magnetosheath show that they contribute very little heat individually \citep{Phan2018}. Due to their statistical dominance, however, they may contribute non-negligibly to heating in aggregate. 

It is important to note that the distribution may be limited by the simulation box size. For fast reconnecting sheets expected to have an aspect ratio of $\sim 10$ \citep{Liu2017,Mbarek2022}, if most of the sheets were $2 d_i$, our distribution would be dominated by structures whose perpendicular length scale was comparable to the coherence length. This motivates a dedicated follow-up study to determine whether the scale distribution remains peaked at electron scales in a larger simulation volume. However, the increase needed may be out of reach currently as achieving comparable scale separation at ion scales as is currently resolved at electron scales would require a simulation volume roughly $\sim 350$ times larger.

These results establish a direct link between kinetic simulations and single-spacecraft observations by evaluating both PVI's ability to detect CS and its capacity to infer their trajectory scales. Resolving these electron-scale structures, that single-spacecraft instruments can struggle to measure directly, requires a fully kinetic and 3D treatment that captures both the kinetic physics setting their thickness and the full geometry of the turbulent cascade. The dominance of sub-ion scale current sheets suggests that turbulent dissipation via reconnection may receive a substantial contribution from electron-scale events, and that solar-wind heating may arise partly from the collective contribution of many intermittent small-scale structures rather than only from rare, large-scale reconnection episodes. Future multi-point missions such as Plasma Observatory and HelioSwarm, which will sample turbulence simultaneously across multiple scales, can test these predictions more rigorously and help determine whether electron-scale dominance is a universal feature of solar wind turbulence.

\acknowledgments
The work of ZD, DS and CCH was supported in part by NSF/DOE Grant PHY-2205991, NSF-FDSS Grant AGS-1936393 and NSF-CAREER Grant AGS-2338131.
LC acknowledges support from NSF grant PHY-2308944 and NASA ATP grant 80NSSC24K1230.
Additionally, the work of ZD, LC and CCH was supported by NSF PHY-2607293. AC acknowledge financial support from the Centre national d’études spatiales (CNES), France (ROR: https://ror.org/04h1h0y33), within the framework of the HelioSwarm and Plasma Observatory space mission.

\bibliographystyle{apsrev4-2}
\bibliography{main}

\end{document}